\documentclass[a4paper,11pt]{article}

\usepackage{jheppub} 
\usepackage[utf8]{inputenc}
\usepackage[skip=6pt,indent]{parskip} 

\usepackage{bm}
\usepackage{comment} 
\usepackage{mathtools}
\usepackage{graphicx}
\usepackage{amsmath,latexsym,amssymb,mathrsfs,ascmac}
\usepackage{multirow}
\usepackage{braket}
\usepackage{color}
\usepackage{enumitem}

\hypersetup{colorlinks,citecolor=TokiwaIro,linkcolor=RuriIro,urlcolor=RuriIro,bookmarksopen,bookmarksnumbered,pdfstartview=FitH,linktocpage}
\definecolor{RuriIro}{rgb}{0.,0.28,0.60}
\definecolor{TokiwaIro}{rgb}{0.,0.39,0.16}

\definecolor{dgreen}{rgb}{0.2,0.51,0.19}

\newcommand{\nn}{\nonumber}

\newcommand{\mc}{\mathcal}
\newcommand{\mr}{\mathrm}

\makeatletter
\gdef\@fpheader{}
\makeatother

\begin{document}

\title{
Causality Constraints on Black Hole Thermodynamics in Nonlinear Electrodynamics}

\date{\today}

\author[a,b,c]{Yoshihiko Abe,}

\author[d,e]{Maxime Médevielle,}

\author[d]{Toshifumi Noumi,}

\author[d]{and Kaho Yoshimura}

\affiliation[a]{Graduate School of Science and Technology, Keio University,\\
Yokohama, Kanagawa 223-8522, Japan}
\affiliation[b]{Keio University Sustainable Quantum Artificial Intelligence Center (KSQAIC),\\
Keio University, Tokyo 108-8345, Japan}
\affiliation[c]{Quantum Computing Center, Keio University,\\
3-14-1 Hiyoshi, Kohoku-ku, Yokohama, Kanagawa, 223-8522, Japan}
\affiliation[d]{Graduate School of Arts and Sciences, University of Tokyo,\\ Komaba, Meguro-ku, Tokyo 153-8902, Japan}
\affiliation[e]{QAT team, DIENS, École Normale Supérieure, PSL University, CNRS, INRIA,\\
45 rue d’Ulm, Paris 75005, France}

\emailAdd{yabe3@keio.jp}
\emailAdd{maxime.medevielle@inria.fr}
\emailAdd{tnoumi@g.ecc.u-tokyo.ac.jp}
\emailAdd{yoshimura-kaho848@g.ecc.u-tokyo.ac.jp}

\abstract{
We study causality constraints on black hole thermodynamics in nonlinear electrodynamics, where the Lagrangian is taken to be an arbitrary function of the electromagnetic field strength tensor. By requiring the absence of superluminal propagation, we show that the mass-to-charge ratio of extremal black holes exhibits a certain monotonicity previously studied in the context of the weak gravity conjecture. Furthermore, under the same condition, we demonstrate that the entropy-to-mass-squared ratio of black holes, which we interpret as an entropy density, decreases monotonically with increasing mass, while keeping the mass-to-charge ratio fixed. This new monotonicity property extends previous studies on the positivity of four-derivative corrections to black hole entropy in the microcanonical ensemble to all orders in nonlinear electrodynamics.
}

\maketitle

\section{Introduction}
Black hole thermodynamics offers a valuable testbed for quantum aspects of gravity. In fact, thought experiments on charged black holes have been used to motivate various Swampland conditions~\cite{Vafa:2005ui} such as the weak gravity conjecture~\cite{Arkani-Hamed:2006emk} (see e.g., Refs.~\cite{Palti:2019pca,vanBeest:2021lhn,Agmon:2022thq,Harlow:2022ich} for review articles). At the same time, research activities toward proof of the conjectured quantum gravity constraints in turn revealed interesting properties of black hole thermodynamics. For instance, the mass-to-charge ratio of extremal black holes turned out to decrease once higher-derivative corrections are taken into account in a wide class of theories~\cite{Natsuume:1994hd, Kats:2006xp,Cheung:2018cwt,Hamada:2018dde,Bellazzini:2019xts, Charles:2019qqt, Jones:2019nev, Loges:2019jzs,Goon:2019faz, Cano:2019oma, Cano:2019ycn,Cremonini:2019wdk, Chen:2020rov, Andriolo:2020lul,Loges:2020trf,  Bobev:2021oku, Arkani-Hamed:2021ajd, Cremonini:2021upd, Aalsma:2021qga,Ma:2021opb,Noumi:2022ybv,Abe:2023anf,Xiao:2025llj,Barbosa:2025smt}. Furthermore, black hole entropy  in the microcanonical ensemble was found to be increased by higher-derivative corrections under similar circumstances~\cite{Cheung:2018cwt,Hamada:2018dde,Loges:2019jzs,Goon:2019faz,Noumi:2022ybv}.

In this paper, motivated by these developments, we study causality constraints on black hole thermodynamics in nonlinear electrodynamics. By requiring the absence of superluminal propagation, we
show that the mass-to-charge ratio of extremal black holes increases monotonically with increasing charge. In addition, under the
same condition, we demonstrate that the entropy-to-mass-squared ratio of black holes, which we interpret as an entropy density, decreases monotonically with increasing mass, while keeping the mass-to-charge ratio fixed. This extends the aforementioned findings in the literature to all orders in nonlinear electrodynamics in a model-independent fashion.

The paper is organized as follows. In section~\ref{sec:BHThermo-NLED}, we review general aspects of nonlinear electrodynamics and summarize the thermodynamics of magnetic black holes. In particular, we write down the mass as a function of the entropy and charge, which carries all the thermodynamic information, reproducing the modified Smarr relation derived in~\cite{Gulin:2017ycu}. In section~\ref{sec:causalityonthermo}, we study causality constraints on thermodynamics of magnetic black holes. As studied in~\cite{Schellstede:2016zue}, absence of superluminal propagation around electromagnetic backgrounds requires a convexity of the Lagrangian function. By combining it with the modified Smarr relation, we derive the monotonicity of the extremal condition and the entropy density. In section~\ref{sec:dyonic}, we extend the analysis to general dyonic black holes by using a dual description after the electromagnetic charge rotation. Discussion of our results is given in section~\ref{sec:conclusion}. Several technical details are collected in the appendices.

\section{Black hole thermodynamics in nonlinear electrodynamics}
\label{sec:BHThermo-NLED}

In this section, we summarize black hole thermodynamics in nonlinear electrodynamics. In particular, we introduce the Smarr-type relation for magnetic black holes~\cite{Smarr:1972kt,Zhang:2016ilt,Gulin:2017ycu,Bokulic:2021dtz}, which serves as the basis for our discussion of causality constraints on black hole thermodynamics.

\subsection{Nonlinear electrodynamics}

We consider four-dimensional nonlinear electrodynamics~\cite{Born:1933qff,Born:1934gh,Boillat:1970gw,osti_4071071,Bialynicki-Birula:1984daz} coupled to Einstein gravity in asymptotically flat spacetime:
\begin{align}
\label{NLaction}
    S=\int d^4x \sqrt{-g} \left[\frac{R}{2}+\mathcal{L}\left(\mathcal{F},\mathcal{G}\right)\right]\,,
\end{align}
where we chose the unit $8\pi G=1$ with the Newton constant $G$. $\mathcal{L}\left(\mathcal{F},\mathcal{G}\right)$ is a function of $\mathcal{F}$ and $\mathcal{G}$ defined by
\begin{align}
\label{eq:mathcalFG}
	\mathcal{F} \coloneqq \frac{1}{4} F_{\mu\nu} F^{\mu\nu}\,,
    \quad
    \mathcal{G} \coloneqq \frac{1}{4} F_{\mu\nu} \tilde{F}^{\mu\nu}
    \quad
    {\rm with}
    \quad
    \tilde{F}^{\mu\nu} :=\frac{1}{2}\epsilon^{\mu\nu\rho\sigma} F_{\rho \sigma}\,,
\end{align}
where $\epsilon^{\mu\nu\rho\sigma}$ is the antisymmetric tensor normalized as $\epsilon^{0123}=-1/\sqrt{-g}$ (see also~\cite[App.~B]{Abe:2023anf} for details of this convention). We assume $\mathcal{L}(0,0)=0$ to set the cosmological constant to zero. This offers an effective theory applicable when electromagnetic fields are strong, but their spacetime-dependence is weak enough. See, e.g.,~\cite{Abe:2023anf} for more on the parameter regime applicable in the black hole context.

The modified Maxwell equations for nonlinear electrodynamics follow from~\eqref{NLaction} as
\begin{align}
\label{eq:modifiedMaxwell}
     \nabla_\mu \frac{\partial \mathcal{L}}{\partial F_{\mu\nu}}
     =0
     \,, \qquad \nabla_\mu \tilde{F}^{\mu\nu}=0~.
\end{align}
To work with them, it is convenient to introduce a two-form field $H_{\mu\nu}$ such that
\begin{align}
\label{def:PL}
    \tilde{H}^{\mu\nu} = 2\frac{\partial \mathcal{L}}{\partial F_{\mu\nu}}=\mathcal{L}_{\mathcal{F}}F^{\mu\nu}+\mathcal{L}_{\mathcal{G}}\tilde{F}^{\mu\nu}~,
\end{align}
where $\mathcal{L}_{\mathcal{F}}$ and $\mathcal{L}_{\mathcal{G}}$ represent the partial derivatives of $\mathcal{L}\left(\mathcal{F},\mathcal{G}\right)$ with respect to $\mathcal{F}$ and $\mathcal{G}$, respectively.
More explicitly, we define
\begin{align}
\label{eq:Pexcitation}
    H_{\mu\nu}\coloneqq -\mathcal{L}_{\mathcal{F}}\tilde{F}_{\mu\nu}+\mathcal{L}_{\mathcal{G}}{F}_{\mu\nu}\,,
\end{align}
in terms of which the modified Maxwell equations are of the Bianchi identity type:
\begin{align}
\label{eq:modifiedMaxwell2}
     \nabla_\mu \tilde{H}^{\mu\nu}
     =0
     \,, \qquad \nabla_\mu \tilde{F}^{\mu\nu}=0~.
\end{align}
Also, the magnetic and electric charges read
\begin{align}
    Q_m=\int_{S^2_\infty}F\,, \qquad Q_e=\int_{S^2_\infty}H\,,
\end{align}
where $S^2_\infty$ is the $S^2$ at the asymptotic infinity.
This manifests that $H_{\mu\nu}$ is the electromagnetic dual of $F_{\mu\nu}$. Indeed, in Maxwell theory $\mathcal{L}=-\mathcal{F}$, we have $H_{\mu\nu}=\widetilde{F}_{\mu\nu}$. 

Now we consider static and spherically symmetric configurations and make the ansatz,
\begin{align}
    F&=F_{tr}(r)dt\wedge dr +F_{\theta\phi}(r) d\theta \wedge d\phi\,,
    \\
    ds^2&=-f(r)dt^2+\frac{1}{f(r)}dr^2+r^2d\theta^2 +r^2\sin^2\theta d\phi^2\,.
\end{align}
Then, from the Bianchi identity of $F_{\mu\nu}$, i.e., the second equation of \eqref{eq:modifiedMaxwell2}, $F_{\theta\phi}$ follows as
\begin{align}
\label{eq:boundary_F}
F_{\theta\phi}&=\frac{\sin\theta}{4\pi}Q_m\, .
\end{align}
Although $F_{tr}$ cannot be fixed in a simple manner, the first equation of \eqref{eq:modifiedMaxwell2} fixes $H_{\theta\phi}$ as
\begin{align}
    \label{eq:boundary_P}
H_{\theta\phi}&=\frac{\sin\theta}{4\pi}Q_e\,.
\end{align}
Substituting~\eqref{eq:boundary_F}--\eqref{eq:boundary_P} into the definition~\eqref{eq:Pexcitation} of $H_{\mu\nu}$, we find
\begin{align}
F_{tr}=\frac{1}{4\pi r^2 \mathcal{L}_{\mathcal{F}}}\left(Q_e-\mathcal{L}_{\mathcal{G}} Q_m\right)\,.
\end{align}
Note that in the Maxwell theory $\mathcal{L}=-\mathcal{F}$, we reproduce $\displaystyle F_{tr}=-\frac{Q_e}{4\pi r^2}$.

To summarize, static and spherically symmetric solutions for the modified Maxwell equations~\eqref{eq:modifiedMaxwell} are given in terms of $Q_e,Q_m,\mathcal{L}_{\mathcal{F}}$ as
\begin{align}
    F=\frac{1}{4\pi r^2 \mathcal{L}_{\mathcal{F}}}\left(Q_e-
    \mathcal{L}_{\mathcal{G}} Q_m\right)dt\wedge dr+\frac{Q_m\sin\theta}{4\pi} d\theta \wedge d\phi~.
\end{align}
Correspondingly, $\mathcal{F}$ and $\mathcal{G}$ read
\begin{align}
    \mathcal{F}
    =\frac{1}{32\pi^2r^4} \left[Q_m^2-\frac{\left(Q_e-\mathcal{L}_{\mathcal{G}}Q_m\right)^2}{\mathcal{L}_{\mathcal{F}}^2}\right]\, ,
    \quad
    \mathcal{G}
    =\frac{1}{16\pi^2  r^4 \mathcal{L}_{\mathcal{F}}}\left(-Q_e+\mathcal{L}_{\mathcal{G}}Q_m\right)Q_m~.
    \label{eq:solution-mcFmcG}
\end{align}

\subsection{Magnetic black holes}
\label{sec:magnetic-BH}

In this subsection, we solve the Einstein equations to construct charged black hole solutions.\footnote{
See, e.g.,~\cite{Pellicer:1969cf,Demianski:1986wx,Breton:2003tk,Breton:2007bza,Kruglov:2017mpj,Chinaglia:2017uqd,Poshteh:2020sgp,Nomura:2020tpc,Nomura:2021efi,Bronnikov:2017xrt,Panahiyan:2018fpb,Wang:2019dzl,Kruglov:2019ybs,Kruglov:2019mvk,Kruglov:2020aqm,Yao:2020ftk,Kruglov:2019okd,Amirabi:2020mzv,Mehdipour:2021ipf,Yang:2023xzv,Toshmatov:2023qbi,Quijada:2023fkc,Magos:2023nnb,Junior:2023ixh,Mignemi:2023uqq,Junior:2024xmm,Vachher:2024ldc,Guzman-Herrera:2024fkg,AraujoFilho:2024xhm,AraujoFilho:2024lsi,Rois:2024iiu,Afshar:2024dhf,Verbin:2024ewl,Hale:2025veb,AraujoFilho:2025hnf,dePaula:2025kif,Tlemissov:2025nnk,Junior:2025izx,Kala:2025kkm,Liu:2025xlr} for earlier works on charged black holes in nonlinear electrodynamics. 
}
We focus on magnetic black holes from here to section~\ref{sec:causalityonthermo}, and discuss extension to dyonic black holes in Sec~\ref{sec:dyonic}. 
 Without electric charge,  $F_{\mu\nu}$ is reduced to
\begin{align}
    F=-\frac{Q_m \mathcal{L}_{\mathcal{G}} }{4\pi r^2 \mathcal{L}_{\mathcal{F}}}dt\wedge dr+\frac{Q_m\sin\theta}{4\pi} d\theta \wedge d\phi\,,
\end{align}
and correspondingly  $\mathcal{F}$ and $\mathcal{G}$ simplify as
\begin{align}
\label{FG_mag}
    \mathcal{F}=\frac{Q_m^2}{32\pi^2r^4} \frac{\mathcal{L}_{\mathcal{F}}^2-\mathcal{L}_{\mathcal{G}}^2}{\mathcal{L}_{\mathcal{F}}^2}\,,
\quad
\mathcal{G}=\frac{Q_m^2\mathcal{L}_{\mathcal{G}}}{16\pi^2 r^4 \mathcal{L}_{\mathcal{F}}}\, .
\end{align}
For visual clarity, we assume $Q_m>0$ without loss of generality.
In the Einstein equations,
\begin{align}
\label{eq:Einstein}
    G_{\mu\nu}=T_{\mu\nu}\,,
\end{align}
nonzero components of the Einstein tensor $G_{\mu\nu}\coloneqq R_{\mu\nu}-\frac{1}{2}g_{\mu\nu}R$ are
\begin{align}
&G^t{}_t=G^r{}_r=\frac{-1+(\theta_r+1)f(r)}{r^2}
\,,
\quad
G^\theta{}_\theta=G^\phi{}_\phi=\frac{\theta_r(\theta_r+1)f(r)}{2r^2}\,,
\end{align}
where $\theta_x=x\partial_x$ is the Euler operator counting the exponent of $x$. On the other hand, the energy-momentum tensor $T_{\mu\nu}$ is of the form,
\begin{align}
\label{eq:EM-tensor}
	T_{\mu\nu} = - \mathcal{L}_{\mathcal{F}} F_{\mu\lambda} F_{\nu}{}^{\lambda} + g_{\mu\nu} \left[ \mathcal{L(\mathcal{F},\mathcal{G})} - \mathcal{G} \mathcal{L}_{\mathcal{G}} \right]\,,
\end{align}
whose nonzero components are
\begin{align}
T^t{}_t=T^r{}_r=\mathcal{L}
\,,
\quad
T^\theta{}_\theta=T^\phi{}_\phi= \left[1-2(\theta_{\mathcal{F}}+\theta_{\mathcal{G}})\right] \mathcal{L}\,. 
\end{align}
With these expressions, the Einstein equations reduce to the following two equations:
\begin{align}    
\label{Einstein_eq1}
\left(\theta_r+1\right)f(r)&=1+ r^2 \mathcal{L}\,,
    \\
\label{Einstein_eq2}
\theta_r\left(\theta_r+1\right)f(r)&=r^2\left[2-4(\theta_{\mathcal{F}}+\theta_{\mathcal{G}})\right] \mathcal{L}\,.
\end{align}
Note that the second equation~\eqref{Einstein_eq2} follows from the first one~\eqref{Einstein_eq1} by using the following relation derived in App.~\ref{app:details}:
\begin{align}
\label{useful_r}
\theta_r\mathcal{L}=-4(\theta_{\mathcal{F}}+\theta_{\mathcal{G}})\mathcal{L}\,.
\end{align}
Therefore, we focus on the first equation~\eqref{Einstein_eq1} in the following analysis.

Using the Einstein equation~\eqref{Einstein_eq1}, we can determine $f(r)$ as
\begin{align}
    f(r)=1- \frac{M}{4\pi r}+\frac{1}{r}\int^r_\infty \frac{dr'}{r'} {r'}^3 \mathcal{L} \left(\mathcal{F},\mathcal{G}\right)\,,
\end{align}
where $\mathcal{F}$, $\mathcal{G}$ are evaluated at the radius $r'$ and the integration constant $M$ is interpreted as the black hole mass. The horizon radius $r_{\rm H}$ is determined by solving $f(r_{\rm H})=0$, in terms of which the black hole mass $M$ reads
\begin{align}
    M=4\pi r_{\rm H}+4\pi \int^{r_{\rm H}}_\infty \frac{dr'}{r'} {r'}^3\mathcal{L}\left(\mathcal{F},\mathcal{G}\right)\,.
\end{align}
For black hole thermodynamics, it is useful to express the mass as a function of the entropy $S=8\pi^2 r_{\rm H}^2$ and the magnetic charge $Q_m$ as 
\begin{align}
\label{eq:mass}
    M(S,Q_m)=\sqrt{2S}+\frac{1}{8\sqrt{2}\pi^2} \int^{S}_\infty \frac{d S'}{S'}  S'^{3/2}\mathcal{L}\left(\mathcal{F}(S',Q_m),\mathcal{G}(S',Q_m)\right) \,,
\end{align}
where $\mathcal{F}$, $\mathcal{G}$ are interpreted as a function of the entropy $S$ and the charge $Q_m$ as
\begin{align}
    \mathcal{F}(S,Q_m)=\frac{2\pi^2Q_m^2}{S^2} \frac{\mathcal{L}_{\mathcal{F}}^2-\mathcal{L}_{\mathcal{G}}^2}{\mathcal{L}_{\mathcal{F}}^2}\,,
\quad
\mathcal{G}(S,Q_m)=\frac{4\pi^2Q_m^2}{S^2 }\frac{\mathcal{L}_{\mathcal{G}}}{\mathcal{L}_{\mathcal{F}}}\, .
\end{align}
Note that  $\mathcal{L}_\mathcal{F}$ and $\mathcal{L}_\mathcal{G}$ here should also be interpreted as a function of $S$ and $Q_m$ under the identification $r=\sqrt{\frac{S}{8\pi^2}}$ of the radius $r$.

\subsection{Thermodynamics}
\label{sec:thermodynamics}

In the previous subsection, we derived the mass formula~\eqref{eq:mass},
\begin{align}
\label{eq:mass2}
    M(S,Q_m)=\sqrt{2S}+\frac{1}{8\sqrt{2}\pi^2} \int^{S}_\infty \frac{d S'}{S'}  S'^{3/2}\mathcal{L} \,,
\end{align}
as a function of the entropy $S$ and the magnetic charge $Q_m$, which serves as a thermodynamic potential that carries all the thermodynamic information. Also, in App.~\ref{app:details}, we derive the following useful formulae that hold for magnetic black holes:
\begin{align}
\label{useful2}
\theta_S\mathcal{L}=-2(\theta_{\mathcal{F}}+\theta_{\mathcal{G}})\mathcal{L}
\,,
\quad
\theta_{Q_m}\mathcal{L}=2(\theta_{\mathcal{F}}+\theta_{\mathcal{G}})\mathcal{L}
\,,
\end{align}
where the first relation is equivalent to the relation~\eqref{useful_r} since $r\propto S^{1/2}$. Eqs.~\eqref{eq:mass2}--\eqref{useful2} and standard thermodynamic relations are all we need in the following analysis.

First, we can use the first law,
\begin{align}
    dM=T dS + \Phi_m dQ_m \,,
\end{align}
to determine the temperature $T$ as
\begin{align}
\label{eq:T-in-NLE}
T&=\left(\frac{\partial M}{\partial S}\right)_{Q_m}=\frac{1}{\sqrt{2S}}+\frac{\sqrt{S}}{8\sqrt{2}\pi^2}\mathcal{L}\left(\mathcal{F},\mathcal{G}\right)\,,
\end{align}
which leads to
\begin{align}
ST&=\sqrt{\frac{S}{2}}+\frac{1}{8\sqrt{2}\pi^2}\int_\infty^S \frac{dS'}{S'} \theta_{S'}\left({S'}^{3/2}\mathcal{L}\right)
\nonumber
\\
\label{ST}
&=\sqrt{\frac{S}{2}}+\frac{1}{8\sqrt{2}\pi^2}\int_\infty^S \frac{dS'}{S'}S^{'3/2}\left[\frac{3}{2}-2\left(\theta_{\mathcal{F}}+\theta_{\mathcal{G}}\right)\right]\mathcal{L}
\,.
\end{align}
Here we used \eqref{useful2} at the second equality. Similarly, the magnetic potential $\Phi_m$ reads
\begin{align}
\Phi_m &=\left(\frac{\partial M}{\partial Q_m}\right)_S=\frac{1}{8\sqrt{2}\pi^2} \int^{S}_\infty \frac{d S'}{S'}  S'^{3/2}\left(\frac{\partial \mathcal{L}}{\partial Q_m}\right)_{S'}\, ,
    \label{eq:Phi-in-NLE}
\end{align}
so that we have
\begin{align}
Q_m\Phi_m &=\frac{1}{8\sqrt{2}\pi^2} \int^{S}_\infty \frac{d S'}{S'}  S'^{3/2}\theta_{Q_m}\mathcal{L}
\nonumber
\\
\label{QPhi}
&=\frac{1}{4\sqrt{2}\pi^2} \int^{S}_\infty \frac{d S'}{S'}  S'^{3/2}\left(\theta_{\mathcal{F}}+\theta_{\mathcal{G}}\right)\mathcal{L}\,,
\end{align}
where we used \eqref{useful2} at the second equality.

Combining \eqref{eq:mass2}, \eqref{ST}, and \eqref{QPhi}, we reproduce the modified Smarr relation for nonlinear electrodynamics derived in \cite{Gulin:2017ycu}:
\begin{align}
\label{eq:modifiedsmarr}
M(S,Q_m)=2ST+Q_m\Phi_m+\frac{1}{4\sqrt{2}\pi^2} \int^{S}_\infty \frac{d S'}{S'}  S'^{3/2}\left(\theta_\mathcal{F}+\theta_{\mathcal{G}}-1\right)\mathcal{L}\,.
\end{align} 
Note that the last term is the correction from the nonlinearity.

\section{Causality constraints on thermodynamics}
\label{sec:causalityonthermo}

As shown in~\cite{Schellstede:2016zue}, absence of superluminal propagation of electromagnetic waves around background electromagnetic fields requires convexity of the Lagrangian $\mathcal{L(\mathcal{F},\mathcal{G})}$:\footnote{
\label{footnote:caveat}
A caveat is needed here: The condition~\eqref{eq:DEC} was derived in the absence of dynamical gravity, so more careful causality analysis is needed to derive rigorous constraints on the present setup. Indeed, in the presence of dynamical gravity, the notion of lightcone is not invariant under field redefinition of the metric and so the subluminal condition is not well defined. However, we study implications of~\eqref{eq:DEC}, having in mind situations where gravitational corrections are negligible. See App.~\ref{app:dynamical_gravity} for more details.
}
\begin{align}
\label{eq:DEC}
\left(\theta_{\mathcal{F}}+\theta_{\mathcal{G}}\right)\mathcal{L}(\mathcal{F},\mathcal{G})\geq \mathcal{L}(\mathcal{F},\mathcal{G})\,.
\end{align}
Note that the causality condition \eqref{eq:DEC} is equivalent to the dominant energy condition~\cite{Bokulic:2021dtz,Russo:2024xnh,dePaula:2024yzy}, which requires that matter propagates along time-like or null geodesics.
See~\cite{Russo:2024xnh} for a detailed discussion on the relation between causality and various energy conditions in nonlinear electrodynamics.
For example, the causality condition~\eqref{eq:DEC} requires that the nonlinear correction in the modified Smarr relation~\eqref{eq:modifiedsmarr} is negative~\cite{Bokulic:2021dtz}:
\begin{align}
\label{eq:Smarrinequality}
    M(S,Q_m)&=2ST+Q_m\Phi-\frac{1}{4\sqrt{2}\pi^2}\int_{S}^\infty \frac{d S'}{S'}  S'^{3/2}\left(\theta_\mathcal{F}+\theta_{\mathcal{G}}-1\right)\mathcal{L}
    \nonumber\\
    &\leq 2ST+Q_m\Phi~.
\end{align}
In the following, we study further implications of causality for black hole thermodynamics.

\subsection{Four-derivative corrections}

We first summarize the known results on four-derivative corrections, based on the thermodynamics introduced in the previous section. For simplicity, let us focus on the parity-even corrections of the form,
\begin{align}
\label{even_4der}
\mathcal{L}\left(\mathcal{F},\mathcal{G}\right)=-\mathcal{F}+\alpha_1\mathcal{F}^2+\alpha_2\mathcal{G}^2\,.
\end{align}
Note that the convexity condition~\eqref{eq:DEC} implies $\alpha_1\geq0$ and $\alpha_2\geq0$, which follow also from consistency of scattering amplitudes as long as gravitational corrections are negligible~\cite{Adams:2006sv}.\footnote{Similarly to the subluminal condition, positivity bounds on scattering amplitudes are modified in gravity theories, accommodating a tiny negativity suppressed by the Planck scale. See also App.~\ref{app:dynamical_gravity} for more details.}
Using~\eqref{useful2}, we obtain the mass formula~\eqref{eq:mass2} in the present setup as
\begin{align}
\label{mass_4der}
M(S,Q_m)=\sqrt{2S}+
\frac{1}{2\sqrt{2}}\frac{Q_m^2}{S^{1/2}}-\frac{\pi^2}{5\sqrt{2}}\frac{\alpha_1Q_m^4}{S^{5/2}}\,.
\end{align}
Note that the effective coupling $\alpha_2$ did not show up because  $\mathcal{G}=\mathcal{L}_\mathcal{G}=0$ for magnetic black holes in the parity symmetric theory. See also~\eqref{FG_mag}.

\paragraph{Extremal condition.}

The mass-to-charge ratio of extremal black holes is of particular interest in the context of the weak gravity conjecture. To evaluate this ratio, it is convenient to write down an explicit formula for the temperature $T$:
\begin{align}
T&=\left(\frac{\partial M}{\partial S}\right)_{Q_m}=\frac{1}{\sqrt{2S}}
-\frac{1}{4\sqrt{2}}\frac{Q_m^2}{S^{3/2}}
+\frac{\pi^2}{2\sqrt{2}}\frac{\alpha_1Q_m^4}{S^{7/2}}
\,.
\end{align}
By solving the zero-temperature condition $T=0$ perturbatively, the entropy $S_{\rm ext}(Q_m)$ of extremal black holes reads
\begin{align}
S_{\rm ext}(Q_m)=\frac{Q_m^2}{4}-8\pi^2\alpha_1+\mathcal{O}(Q_m^{-2})\,.
\end{align}
Substituting this back into the mass formula~\eqref{eq:mass2} gives the mass of extremal black holes, $M_{\rm ext}(Q_m)=M(S_{\rm ext}(Q_m),Q_m)$, from which the mass-to-charge ratio reads
\begin{align}
\mu_{\rm ext}(Q_m)\coloneqq \frac{M_{\rm ext}(Q_m)}{Q_m}=\sqrt{2}-\frac{16\sqrt{2}\pi^2}{5}\frac{\alpha_1}{Q_m^2}+\mathcal{O}(Q_m^{-4})\,.
\end{align}
This shows that four-derivative corrections to the mass-to-charge ratio of extremal black holes are negative as long as the convexity condition~\eqref{eq:DEC} is satisfied, i.e., $\alpha_1>0$, as expected by the weak gravity conjecture~\cite{Kats:2006xp}.

\paragraph{Entropy correction.}

Another point of interest is the entropy correction in the microcanonical ensemble. To evaluate it, we solve the mass formula~\eqref{mass_4der} perturbatively for entropy as
\begin{align}
\label{S_4der}
S(M,Q_m)=S_0
+\frac{4\pi^2}{5}\frac{\alpha_1Q_m^4}{S_0(4S_0-Q_m^2)}
+\cdots
\,,
\end{align}
where the dots stand for higher order corrections and $S_0$ is the entropy at the leading order given by 
\begin{align}
S_0=\frac{1}{8} \left(\sqrt{M^2-2 Q_m^2}+M\right)^2\,.
\end{align}
For generic black holes $4S_0> Q_m^2$, the entropy correction turns out to be positive as long as the convexity condition~\eqref{eq:DEC} is satisfied, i.e., $\alpha_1>0$~\cite{Cheung:2018cwt}. For black holes with $Q_m=M/\sqrt{2}$, the expression~\eqref{S_4der} is not applicable, but it is known to be positive under the same condition~\cite{Hamada:2018dde}.

\subsection{Full nonlinear electrodynamics}

\paragraph{Formulating the problem.}

As we have just reviewed, the convexity conditions~\eqref{eq:DEC} fix the sign of four-derivative corrections to the extremal condition and the entropy in the microcanonical ensemble. Schematically, we write
\begin{align}
\Delta \mu_{\rm ext}(Q_m)
<0
\,,
\quad
\Delta S(M,Q_m)>0\,.
\end{align}
Now we would like to go beyond the leading-order corrections and study the full nonlinear electrodynamics. In the context of the extremal condition, this extension was formulated in terms of the scale-dependence of the mass-to-charge ratio~\cite{Abe:2023anf}. For instance, in the parity-even four-derivative EFT~\eqref{even_4der}, we have 
\begin{align}
\frac{d\mu_{\rm ext}(Q_m)}{dQ_m}=\frac{32\sqrt{2}\pi^2}{5}\frac{\alpha_1}{Q_m^3}
+\mathcal{O}(Q_m^{-5})\,,
\end{align}
which is fixed to be positive by the convexity condition~\eqref{eq:DEC} as long as $Q_m$ is large enough and the four-derivative approximation works. See also~\cite{Abe:2023anf} for its extension to the Euler-Heisenberg and DBI effective theories. One advantage of this formulation is that the condition is defined in terms of the scale dependence in a single theory without referring to the (two-derivative) Einstein-Maxwell theory.

To formulate the problem of entropy in a similar fashion, we define what we call the {\it entropy density} $s_\mu(M)$ for a fixed mass-to-charge ratio $\mu$ as a function of the mass:
\begin{align}
s_\mu(M):=\frac{S\left(M,\frac{M}{\mu}\right)}{M^2}\,.
\end{align}
Recall that for a fixed $\mu$, the length scale of the black hole is controlled by the mass $M$. Also, we define the entropy density by dividing the mass squared, rather than the mass cubed, reflecting the holographic nature of the black hole entropy. For instance, in the parity-even four-derivative EFT~\eqref{even_4der}, the entropy density reads
\begin{align}
s_\mu(M)=s_0+\frac{4\pi^2}{5}\frac{\alpha_1}{s_0\mu^2(4s_0\mu^2-1)M^2}+\mathcal{O}(M^{-4})\,,
\end{align}
where $s_0$ is the entropy density in the Einstein-Maxwell theory that is scale independent:
\begin{align}
s_0=\frac{1}{8}\left(\sqrt{1-2\mu^{-2}}+1\right)^2\,.
\end{align}
Correspondingly, the scale dependence of the entropy density reads
\begin{align}
\frac{d s_\mu(M)}{d M}=-\frac{8\pi^2}{5}\frac{\alpha_1}{s_0\mu^2(4s_0\mu^2-1)M^3}+\mathcal{O}(M^{-5})\,,
\end{align}
which is fixed to be negative by the convexity condition~\eqref{eq:DEC} as long as $M$ is large enough and the four-derivative approximation works.

In the rest of this subsection, we show that the convexity condition~\eqref{eq:DEC} requires
\begin{align}
\frac{d\mu_{\rm ext}(Q_m)}{dQ_m}\geq 0\,,
\quad
\frac{ds_\mu(M)}{dM}\leq0\,,
\end{align}
in the full nonlinear electrodynamics beyond the leading order corrections. In particular, we do not assume parity invariance in contrast to the earlier four-derivative analysis.

\paragraph{Extremal condition.}

From the standard thermodynamics relations, the scale dependence of the mass-to-charge ratio $\mu=M/Q_m$ for fixed temperature $T$ reads
\begin{align}
\nonumber
\left(\frac{\partial \mu}{\partial Q_m}\right)_T
&=Q_m^{-1}\left(\frac{\partial M}{\partial Q_m}\right)_T-\frac{M}{Q_m^2}
\\
&=\frac{Q_m\Phi_m-M}{Q_m^2}+TQ_m^{-1}\left(\frac{\partial S}{\partial Q_m}\right)_T\,,
\end{align}
whose last term vanishes at $T=0$ as long as the entropy of extremal black holes is a smooth function of $Q_m$. Then, using the modified Smarr relation~\eqref{eq:Smarrinequality} for $T=0$, we find
\begin{align}
    \frac{d\mu_{\mr{ext}}(Q_m)}{d Q_m} = \frac{1}{4\sqrt{2}\pi^2 Q_m^2} \int_S^\infty \frac{d S'}{S'} S'^{3/2} (\theta_{\mc{F}} + \theta_{\mc{G}} -1) \mc{L} \geq 0\,,
\end{align}
where we used the convexity condition~\eqref{eq:DEC} at the inequality.

\paragraph{Entropy density.}

First, the scale dependence of the entropy for fixed mass-to-charge ratio $\mu$ follows from the standard thermodynamic relations as
\begin{align}
\left(\frac{\partial S}{\partial M}\right)_{\mu}
=\left(\frac{\partial S}{\partial M}\right)_{Q_m}+\left(\frac{\partial S}{\partial Q_m}\right)_M\left(\frac{\partial Q_m}{\partial M}\right)_\mu
=\frac{M-Q_m\Phi_m}{TM}
\,,
\end{align}
where we used $\left(\frac{\partial Q_m}{\partial M}\right)_\mu=\mu^{-1}=\frac{Q_M}{M}$ in particular. Then, the scale dependence of the entropy density $s_\mu$ for fixed mass-to-charge ratio $\mu$ reads
\begin{align}
\frac{ds_\mu(M)}{dM}&=M^{-2}\left(\frac{\partial S}{\partial M}\right)_{\mu}-2\frac{S}{M^3}
=\frac{M-Q_m\Phi_m-2ST}{TM^3}\,.
\end{align}
Using the modified Smarr relation~\eqref{eq:Smarrinequality}, we conclude that
\begin{align}
    \frac{d s_\mu(M)}{dM} =- \frac{1}{4 \sqrt{2} \pi^2 TM^3} \int_S^\infty \frac{dS'}{S'} S'^{3/2} (\theta_\mc{F} + \theta_\mc{G} -1 ) \mc{L} \leq 0\,.
\end{align}

\section{Extension to dyonic black holes}
\label{sec:dyonic}

For completeness, we extend the analysis to dyonic black holes. For this, we first introduce the electromagnetic rotation in nonlinear electrodynamics in section~\ref{subsec:rotation}. We then apply it to black hole thermodynamics and map the analysis of dyonic black holes to that of magnetic black holes in the dual theory.  

\subsection{Electromagnetic rotation in nonlinear electrodynamics}
\label{subsec:rotation}

\paragraph{Electromagnetic rotation.}

First, let us introduce a two-form field $P_{\mu\nu}$ defined by
\begin{align}
\label{def:rotation}
P_{\mu\nu}:=\cos\alpha\,F_{\mu\nu}+\sin\alpha\,H_{\mu\nu}\,,
\end{align}
where $\alpha$ plays a role of the rotation angle on the electromagnetic charge plane. Similarly to \eqref{eq:mathcalFG}, we define
\begin{align}
\label{eq:mathcalPQ}
	\mathcal{P} \coloneqq \frac{1}{4} P_{\mu\nu} P^{\mu\nu}\,,
    \quad
    \mathcal{Q} \coloneqq \frac{1}{4} P_{\mu\nu} \tilde{P}^{\mu\nu}\,.
\end{align}
In this language, the dual Lagrangian $\bar{\mathcal{L}}(\mathcal{P},\mathcal{Q})$ after the electromagnetic rotation is defined such that
\begin{align}
\label{def:barL}
\bar{\mathcal{L}}(\mathcal{P},\mathcal{Q})=\mathcal{L}(\mathcal{F},\mathcal{G})+ \frac{\cos\alpha}{4\sin\alpha}\tilde{F}^{\mu\nu}F_{\mu\nu}+ \frac{\cos\alpha}{4\sin\alpha}\tilde{P}^{\mu\nu}P_{\mu\nu} -\frac{1}{2\sin\alpha}\tilde{F}^{\mu\nu}P_{\mu\nu}\,.
\end{align}
Indeed, if we define a two-form field $W_{\mu\nu}$ in analogy with \eqref{def:PL} such that
\begin{align}
\tilde{W}^{\mu\nu}
&=2\frac{\partial \bar{\mathcal{L}}(\mathcal{P},\mathcal{Q})}{\partial P_{\mu\nu}}
\,,
\end{align}
we find
\begin{align}
\label{eq:relationWFP}
\sin\alpha\,\tilde{W}^{\mu\nu}
&=
\left[\sin\alpha\,\tilde{H}^{\rho\sigma}+\cos\alpha\,\tilde{F}^{\rho\sigma}-\tilde{P}^{\rho\sigma}\right]\frac{\partial F_{\rho\sigma}}{\partial P_{\mu\nu}}
+\cos\alpha \,\tilde{P}^{\mu\nu}-\tilde{F}^{\mu\nu}
\nonumber
\\
&=\cos\alpha \,\tilde{P}^{\mu\nu}-\tilde{F}^{\mu\nu}\,,
\end{align}
where we used the definition~\eqref{def:rotation} of $P_{\mu\nu}$ at the second equality. This gives the following relation  between $P_{\mu\nu},W_{\mu\nu}$ and $F_{\mu\nu},H_{\mu\nu}$ with a rotation angle $\alpha$:
\begin{align}
\label{rotation-relation}
P_{\mu\nu}=\cos\alpha\,F_{\mu\nu}+\sin\alpha\,H_{\mu\nu}\,,
\quad
W_{\mu\nu}=\cos\alpha\,H_{\mu\nu}-\sin\alpha\,F_{\mu\nu}\,.
\end{align}
Note that the modified Maxwell equations~\eqref{eq:modifiedMaxwell2} in the original theory guarantee those in the dual:
\begin{align}
\label{eq:modifiedMaxwell_dual}
     \nabla_\mu \tilde{W}^{\mu\nu}
     =0
     \,, \qquad \nabla_\mu \tilde{P}^{\mu\nu}=0~.
\end{align}
As designed, the corresponding electromagnetic charges $\bar{Q}_m,\bar{Q}_e$ defined by
\begin{align}
\bar{Q}_m=\int_{S^2_\infty}P\,, \qquad \bar{Q}_e=\int_{S^2_\infty}W
\end{align}
are related to the original ones as 
\begin{align}
\bar{Q}_m=\cos\alpha\, Q_m+\sin\alpha \,Q_e
\,,
\quad
\bar{Q}_e=\cos\alpha \,Q_e-\sin\alpha\, Q_m\,.
\end{align}

\paragraph{Convexity condition.}

Next, we show that the following relation holds identically:
\begin{align}
\label{convexity_preserved}
(\theta_{\mathcal{F}}+\theta_{\mathcal{G}}-1)\mathcal{L}=(\theta_{\mathcal{P}}+\theta_{\mathcal{Q}}-1)\bar{\mathcal{L}}\,,
\end{align}
which implies that the convexity condition~\eqref{eq:DEC} is preserved under the electromagnetic rotation in particular. To show this identity, it is convenient to note the relations,
\begin{align}
(\theta_{\mathcal{F}}+\theta_{\mathcal{G}})\mathcal{L}=\frac{1}{4}F_{\mu\nu}\tilde{H}^{\mu\nu}\,,
\qquad
(\theta_{\mathcal{P}}+\theta_{\mathcal{Q}})\bar{\mathcal{L}}=\frac{1}{4}P_{\mu\nu}\tilde{W}^{\mu\nu}\,,
\end{align}
that follow from the definition of $H_{\mu\nu},W_{\mu\nu}$. See, e.g.,~\eqref{eq:Pexcitation}. Using these relations, we can prove the identity~\eqref{convexity_preserved} as
\begin{align}
&(\theta_{\mathcal{F}}+\theta_{\mathcal{G}}-1)\mathcal{L}-
(\theta_{\mathcal{P}}+\theta_{\mathcal{Q}}-1)\bar{\mathcal{L}}
\nonumber
\\
&=\frac{
(\cos\alpha\,\tilde{F}^{\mu\nu}+\sin\alpha\,\tilde{H}^{\mu\nu}-\tilde{P}^{\mu\nu})F_{\mu\nu}
-(\cos\alpha\,\tilde{P}^{\mu\nu}-\sin\alpha\,\tilde{W}^{\mu\nu}-\tilde{F}^{\mu\nu})P_{\mu\nu}
}{4\sin\alpha}
\nonumber
\\
&=0\,,
\end{align}
where we used the definition~\eqref{def:barL} of $\bar{\mathcal{L}}$ at the first equality and the relations~\eqref{eq:relationWFP}  and~\eqref{rotation-relation} at the second equality.

\paragraph{Energy-momentum tensor.}

Finally, we introduce the dual energy-momentum tensor defined by
\begin{align}
\bar{T}_{\mu\nu} = - \bar{\mathcal{L}}_{\mathcal{P}} P_{\mu\lambda} P_{\nu}{}^{\lambda} + g_{\mu\nu} \left[ \mathcal{\bar{L}(\mathcal{P},\mathcal{Q})} - \mathcal{Q} \bar{\mathcal{L}}_{\mathcal{Q}} \right]\,,
\end{align}
and demonstrate that it coincides with the original energy-momentum tensor, i.e.,
\begin{align}
\bar{T}_{\mu\nu}=T_{\mu\nu}\,.
\end{align}
For this, it is convenient to notice the relation,
\begin{align}
    \mc{L}_\mc{F} F^{\mu\nu} =\frac{1}{\sin\alpha}\tilde{P}^{\mu\nu} - \left(\frac{\cos\alpha}{\sin\alpha} + \mc{L}_\mc{G}\right) \tilde{F}^{\mu\nu}\,,
\end{align}
which follows from~\eqref{def:PL} and the definition~\eqref{def:rotation} of $P_{\mu\nu}$. Using this, we can reformulate the energy-momentum tensor $T_{\mu\nu}$ as
\begin{align}
T^\mu{}_\nu
&= - \left[
        \frac{1}{\sin\alpha} \tilde{P}^{\mu\rho} - \left(\frac{\cos\alpha}{\sin\alpha}+ \mc{L}_\mc{G}\right) \tilde{F}^{\mu\rho}
    \right] F_{\nu\rho} + \delta^\mu_\nu (\mc{L} - \mc{L}_\mc{G} \mc{G})
    \nn \\
    &= -\frac{1}{\sin\alpha} \tilde{P}^{\mu\rho} F_{\nu\rho} + \delta^\mu_\nu \left(\mc{L} + \frac{\cos\alpha}{4\sin\alpha}F_{\mu\nu}\tilde{F}^{\mu\nu}\right) \,.
\end{align}
where at the second equality we used the following identity that holds for general two-form fields $X_{\mu\nu},Y_{\mu\nu}$ in four dimensions:
\begin{align}
\label{eq:twoform}
X^{\mu\rho }Y_{\nu\rho}
-\tilde{Y}^{\mu\rho }\tilde{X}_{\nu\rho}
=\frac{1}{2}\delta^\mu_\nu X_{\alpha\beta}
Y^{\alpha\beta}\,.
\end{align}
Similarly, we have
\begin{align}
\bar{T}^\mu{}_\nu
    &= \frac{1}{\sin\alpha} \tilde{F}^{\mu\rho} P_{\nu\rho} + \delta^\mu_\nu \left(\bar{\mc{L}} - \frac{\cos\alpha}{4\sin\alpha}P_{\mu\nu}\tilde{P}^{\mu\nu}\right) \,.
\end{align}
Using the definition~\eqref{def:barL} of $\bar{\mathcal{L}}$, we find
\begin{align}
T^\mu{}_\nu-\bar{T}^\mu{}_\nu
=\frac{1}{\sin\alpha}
\left(
\frac{1}{2}\delta^\mu_\nu \tilde{F}^{\alpha\beta}P_{\alpha\beta}
-\tilde{P}^{\mu\rho}F_{\nu\rho}-\tilde{F}^{\mu\rho}P_{\nu\rho}
\right)=0\,,
\end{align}
where at the second equality we used \eqref{eq:twoform}. 

\subsection{Causality constraints on thermodynamics of dyonic black holes}

\paragraph{Dyonic black holes.}

Now we apply the electromagnetic rotation introduced in the previous subsection to black holes. Consider a dyonic black hole with electromagnetic charges,
\begin{align}
(Q_m,Q_e)=(Q\cos\alpha,Q\sin\alpha)\,,
\end{align}
where $Q>0$ and $\alpha$ are the amplitude and the angle of the charge vector respectively. By performing the electromagnetic rotation~\eqref{rotation-relation} with the angle $\alpha$, we can map the dyonic black hole to the magnetic black hole in the dual theory described by the Lagrangian $\bar{\mathcal{L}}$:
\begin{align}
(\bar{Q}_m,\bar{Q}_e)=(Q,0)\,.
\end{align}
Since the energy-momentum tensor is invariant under the charge rotation, the black hole solution satisfies
\begin{align}
M(S,Q;\alpha)=\sqrt{2S}+\frac{1}{8\sqrt{2}\pi^2} \int^{S}_\infty \frac{d S'}{S'}  S'^{3/2}\bar{\mathcal{L}} \,,
\end{align}
where $M$ and $S$ are the mass and entropy, respectively.

\paragraph{Thermodynamics.}

Then, the thermodynamics can be studied in the same manner as we did in section~\ref{sec:thermodynamics}. First, by recycling the results there, the product $ST$ is given by
\begin{align}
ST&=\sqrt{\frac{S}{2}}+\frac{1}{8\sqrt{2}\pi^2}\int_\infty^S \frac{dS'}{S'}S^{'3/2}\left[\frac{3}{2}-2\left(\theta_{\mathcal{P}}+\theta_{\mathcal{Q}}\right)\right]\bar{\mathcal{L}}
\,.
\end{align}
Defining the potential along the $\alpha$ direction in the charge space by
\begin{align}
\Phi:=\left(\frac{\partial M}{\partial Q}\right)_{S,\alpha}\,,
\end{align}
we have
\begin{align}
Q\Phi
&=\frac{1}{4\sqrt{2}\pi^2} \int^{S}_\infty \frac{d S'}{S'}  S'^{3/2}\left(\theta_{\mathcal{P}}+\theta_{\mathcal{Q}}\right)\bar{\mathcal{L}}\,.
\end{align}
In this language, the magnetic and electric potentials read
\begin{align}
\Phi_m=\cos\alpha\,\Phi\,,
\quad
\Phi_e=\sin\alpha\,\Phi\,,
\end{align}
which implies
\begin{align}
\label{QPhiem}
Q\Phi=Q_m\Phi_m+Q_e\Phi_e\,.
\end{align}
Combining the above results, we reproduce the modified Smarr relation derived in \cite{Gulin:2017ycu}:
\begin{align}
\nonumber
M&=2ST+Q\Phi+\frac{1}{4\sqrt{2}\pi^2} \int^{S}_\infty \frac{d S'}{S'}  S'^{3/2}\left(\theta_\mathcal{P}+\theta_{\mathcal{Q}}-1\right)\bar{\mathcal{L}}
\\
\label{Smarr-dyonic}
&=2ST+Q_m\Phi_m+Q_e\Phi_e+\frac{1}{4\sqrt{2}\pi^2} \int^{S}_\infty \frac{d S'}{S'}  S'^{3/2}\left(\theta_\mathcal{F}+\theta_{\mathcal{G}}-1\right)\mathcal{L}\,,
\end{align}
where we used~\eqref{convexity_preserved} and~\eqref{QPhiem} at the second equality. Notice that the modified Smarr relation~\eqref{Smarr-dyonic} is invariant under the charge rotation, but each of $M$, $ST$, and $Q\Phi$ are not invariant unless the Lagrangian is invariant, i.e., $\bar{\mathcal{L}}=\mathcal{L}$.

\paragraph{Causality constraints.}

Causality constraints can also be studied in the same manner as before. For dyonic black holes, we denote the mass-to-charge ratio of extremal black holes of the charge $(Q_m,Q_e)=(Q\cos\alpha,Q\sin\alpha)$ by $\mu_{\rm ext}(Q,\alpha)$ and the entropy-to-mass-squared ratio (entropy density) for fixed $\mu=M/Q$ and $\alpha$ by $s_{\mu,\alpha}(M)$. Then, similarly to the magnetic case, we find
\begin{align}
\frac{\partial\mu_{\mr{ext}}(Q,\alpha)}{\partial Q} &= \frac{1}{4\sqrt{2}\pi^2 Q^2} \int_S^\infty \frac{d S'}{S'} S'^{3/2} (\theta_{\mc{F}} + \theta_{\mc{G}} -1) \mc{L} \geq 0\,,
\\
    \frac{d s_{\mu,\alpha}(M)}{dM} &=- \frac{1}{4 \sqrt{2} \pi^2 TM^3} \int_S^\infty \frac{dS'}{S'} S'^{3/2} (\theta_\mc{F} + \theta_\mc{G} -1 ) \mc{L} \leq 0\,.
\end{align}
This extends our previous analysis of magnetic black holes to dyonic black holes.

\section{Discussion}
\label{sec:conclusion}

In this paper, we studied causality constraints on black hole thermodynamics in nonlinear electrodynamics. In particular, we demonstrated that causality implies a new monotonicity property of the entropy-to-mass-squared ratio that we interpret as an entropy density, in addition to the previously discussed monotonicity of the extremal condition. This extends previous findings on higher-derivative corrections discovered in the context of the weak gravity conjecture to all orders in nonlinear electrodynamics. Technically, the modified Smarr relation was useful for the thermodynamic analysis and the electromagnetic rotation significantly simplified the analysis of dyonic black holes.

While the present paper focused on asymptotically flat black holes, extension to the nonzero cosmological constant case is of great interest for various reasons: First, extension to anti-de Sitter (AdS) black holes and its dual CFT interpretation will be useful to support our interpretation of the entropy-to-mass-squared ratio as an entropy density. The dual CFT perspective would also be useful for deriving monotonicity properties in more general grounds beyond the nonlinear electrodynamics. This direction would also sharpen the weak gravity conjecture in AdS, whose formulation is not fully understood yet (see, e.g.,~\cite{Nakayama:2015hga,Montero:2016tif,Crisford:2017gsb,Aharony:2021mpc,Lin:2025wfe}). Similarly, extension to de Sitter (dS) black holes would be useful for better understanding of the weak gravity conjecture in dS and the Festina Lente bound~\cite{Montero:2019ekk}. We hope to revisit these issues in the near future.

\section*{Acknowledgment}
\noindent
The work of YA is supported by the Center of Innovations for Sustainable Quantum AI (JST Grant Number JPMJPF2221). TN is supported in part by JSPS KAKENHI Grant No. JP22H01220 and MEXT KAKENHI Grant No. JP21H05184 and No. JP23H04007. MM is supported by JSPS Postdoctoral Fellowships for Research in Japan
(Standard) No. P23774 and by the French National Research Agency (ANR) under Grant ANR-24-CE92-0076
(BoLaCo). KY is supported by JST SPRING, Grant Number JPMJSP2108 and by a research encouragement grant from the Iwanami Fūjukai. 

\appendix

\section{Causality and dynamical gravity}
\label{app:dynamical_gravity}

As we mentioned in the footnote~\ref{footnote:caveat}, the notion of lightcone becomes ambiguous in the presence of dynamical gravity. For example, consider the following redefinition of the metric:
\begin{align}
    g_{\mu\nu} \rightarrow g_{\mu\nu}+ A(\mathcal{F},\mathcal{G}) F_\mu{}^\rho F_{\nu\rho}\,,
\end{align}
where  $A(\mathcal{F},\mathcal{G})$ parameterizes the disformal transformation. Evidently, the lightcone is not invariant under this redifinition if the electromagnetic field $F_{\mu\nu}$ acquires a nonzero background. Accordingly, the subluminality of fluctuations around the background leads to a frame-dependent constraint on the EFT and it is not clear in which frame and under what conditions we may require subluminality.

To elaborate on this issue, let us first consider the parity-even correction at the leading order. By perturbative field redefinition, which is equivalent to using the leading order equations of motion, one can always eliminate operators that contain $R_{\mu\nu}$ or $\nabla_\mu F^{\mu\nu}$. As a result, a general form of four-derivative corrections reads
\begin{align}
\label{4d_EFT}
S=\int d^4x\sqrt{-g}
\left[\frac{1}{2}M_{\rm Pl}^2R
-\mathcal{F}+\alpha_1\mathcal{F}^2+\alpha_2\mathcal{G}^2
+\beta W_{\mu\nu\rho\sigma}F^{\mu\nu}F^{\rho\sigma}
\right]\,,
\end{align}
where $M_{\rm Pl}=(8\pi G)^{-1/2}$ is the reduced Planck mass and $W_{\mu\nu\rho\sigma}$ is the Weyl tensor. Also, we ignored the Gauss-Bonnet term since it does not affect propagation speed. Note that the Weyl tensor is directly related to the spin-$2$ graviton mode, and hence is a natural ingredient of the effective action after removing redundancy of the field redefinition. In this frame, subluminality of photons in electromagnetic backgrounds implies $\alpha_1\geq0$ and $\alpha_2\geq0$, so that the question is how much one can bound~$\alpha_i$.

\medskip
Remarkably, frame-independent constraints on the gravitational EFT~\eqref{4d_EFT} have been explored, extending the scattering amplitude positivity of non-gravitational EFTs~\cite{Adams:2006sv}. Note that scattering amplitudes are invariant under field redefinition and hence are free from the issue of frame-dependence. A lesson there is that bounds on four-derivative operators in gravitational EFTs are swamped by the $t$-channel graviton pole and a tiny negativity cannot be excluded in contrast to the non-gravitational case (see, e.g.,~\cite{Hamada:2018dde,Alberte:2020jsk,Tokuda:2020mlf,Caron-Huot:2021rmr,Alberte:2021dnj}). In the current setup, the bounds are schematically of the form,
\begin{align}
\label{app_positivity}
\alpha_1\geq \frac{4|\beta|}{M_{\rm Pl}^2} -\frac{C_1}{\Lambda^2M_{\rm Pl}^2}
\,,
\quad
\alpha_2\geq -\frac{C_2}{\Lambda^2M_{\rm Pl}^2}\,,
\end{align}
where $C_i$ are numerical factors and $\Lambda$ is a UV scale that cannot be determined within the low-energy EFT. For example, if the UV theory is the gravitational QED, $\Lambda$ corresponds to the electron mass. This shows that positivity of $\alpha_i$ holds as long as the Planck suppressed terms on the right hand side are negligible, but a tiny negativity cannot be ruled out at least in this analysis. Then, we can translate the approximate positivity bound~\eqref{app_positivity} into {\it an approximate subluminal condition in the frame~\eqref{4d_EFT}}:
\begin{align}
\label{app_p}
v\leq 1+C\frac{\Lambda^2}{M_{\rm Pl}^2}\frac{F^2}{\Lambda^4}\,,
\end{align}
where $v$ is the photon speed in electromagnetic backgrounds, $F$ is strength of the background electromagnetic fields, and $C$ is a positive constant. We emphasize that the amplitude positivity is useful to avoid the issue of frame-dependence, but the exact subluminal condition cannot be obtained even in this approach in the presence of dynamical gravity.

Unfortunately, extension of the discussion to the full nonlinear electrodynamics is not straightforward, essentially because the amplitude positivity of higher-point amplitudes and that in the Lorentz symmetry broken phase have not been well established yet. The best we can do for now is to extrapolate the discussion of four-derivative corrections to the full order on dimensional grounds.
In the context of nonlinear electrodynamics, we are interested in the situation where the electromagnetic fields are not necessarily small, but its inhomogeneity and spacetime curvature are small enough compared to the UV cutoff scale $\Lambda$. Then, we regard the electromagnetic field $F_{\mu\nu}$ as the $0$-th order in derivatives and expand the effective action as
\begin{align}
S=\int d^4x\sqrt{-g}
\left[
\frac{1}{2}M_{\rm Pl}^2R
+\mathcal{L}\left(\mathcal{F},\mathcal{G}\right)
+\mathcal{O}(W, \nabla^2)
\right]\,,
\end{align}
where the last term denotes higher derivative terms that contain at least one Weyl tensor or two derivatives acting on $F_{\mu\nu}$. This ansatz can be thought of as a nonlinear extension of the basis~\eqref{4d_EFT} with a negligible $\beta$ term. On dimensional grounds, we estimate that a nonlinear extension of the approximate subluminal condition~\eqref{app_p} will be schematically of the form,
\begin{align}
v\leq 1+\frac{\Lambda^2}{M_{\rm Pl}^2}f(F^2/\Lambda^4)\,.
\end{align}
Here $f(F^2/\Lambda^4)$ is a dimensionless positive function of $F^2/\Lambda^4$ and its concrete form will depend on details of the UV theory behind. The prefactor $\Lambda^2/M_{\rm Pl}^2$ indicates that the gravitational correction is parametrically suppressed provided that the UV cutoff is well below the Planck scale $\Lambda\ll M_{\rm Pl}$.
Based on this dimensional analysis, we postulate that the  subluminal condition is applicable at least as long as the Planck suppressed gravitational corrections are negligible and study its implication in the main text.

\section{Derivation of~\eqref{useful_r} and~\eqref{useful2}}
\label{app:details}

This appendix provides derivation of~\eqref{useful_r} and~\eqref{useful2}.
Let us consider the following field strength as a solution to the modified Maxwell equation \eqref{eq:modifiedMaxwell}:
\begin{align}
    F = - \frac{Q_m \mc{L}_\mc{G}}{4 \pi r^2 \mc{L}_\mc{F}} dt \wedge dr + \frac{Q_m \sin \theta}{4\pi} d\theta \wedge d \phi\,,
\end{align}
which gives 
\begin{align}
    \mc{F} = \frac{Q_m^2(\mc{L}_\mc{F}^2 - \mc{L}_\mc{G}^2)}{32 \pi^2 r^4 \mc{L}_\mc{F}^2}\,,
    \qquad 
    \mc{G} = \frac{Q_m^2 \mc{L}_\mc{G}}{16 \pi^2 r^4 \mc{L}_\mc{F}}\,.
\end{align}
From these expressions, the action of $\theta_\mc{F} + \theta_\mc{G}$ on the Lagrangian reads 
\begin{align}
    (\theta_\mc{F} + \theta_\mc{G}) \mc{L} = \mc{F} \mc{L}_\mc{F} + \mc{G} \mc{L}_\mc{G} = \frac{Q_m^2(\mc{L}_\mc{F}^2 + \mc{L}_\mc{G}^2)}{32 \pi^2 r^4 \mc{L}_\mc{F}}\,.
    \label{eq:theta-FG-Lag}
\end{align}
Next, we evaluate the action of $\theta_{r}$ and $\theta_{Q_m}$ on the Lagrangian with respect to the radial coordinate $r$ and the magnetic charge $Q_m$.
After direct calculations, we obtain
\begin{align}
   \theta_r\mathcal{L}&= \mc{L}_\mc{F}\, \theta_r \mc{F} + \mc{L}_\mc{G} \,\theta_r\mc{G}
    \nn \\
    &=\mc{L}_\mc{F} \Biggl[
        - \frac{Q_m^2}{8\pi^2 r^4} \biggl( 1 - \frac{\mc{L}_\mc{G}^2}{\mc{L}_\mc{F}^2} \biggr) - \frac{\mc{L}_\mc{G} Q_m}{16 \pi^2 r^4 \mc{L}_\mc{F}^2} \biggl(
            - \frac{\mc{L}_\mc{G}Q_m}{\mc{L}_\mc{F}}\,
            \theta_r\mc{L}_\mc{F} + Q_m \,\theta_r \mc{L}_\mc{G}
        \biggr)
    \Biggr]
    \nn \\
    & \quad 
    + \mc{L}_\mc{G} \Biggl[
        - \frac{Q_m^2 \mc{L}_\mc{G}}{4 \pi^2 r^4 \mc{L}_\mc{F}} + \frac{Q_m}{16 \pi^2 r^4 \mc{L}_\mc{F}} \biggl(
            - \frac{\mc{L}_\mc{G} Q_m}{\mc{L}_\mc{F}}\,
        \theta_r\mc{L}_\mc{F}+ Q_m \,\theta_r \mc{L}_\mc{G}
        \biggr)
    \Biggr]
    \nn \\
    &= - \frac{Q_m^2(\mc{L}_\mc{F}^2 + \mc{L}_\mc{G}^2)}{8 \pi^2 r^4 \mc{L}_\mc{F}}
    \,,
    \label{eq:delLdelr-app}
\end{align}
\begin{align}
    \theta_{Q_m} \mc{L} &= \mc{L}_\mc{F} \,\theta_{Q_m} \mc{F} + \mc{L}_\mc{G} \,\theta_{Q_m} \mc{G}
    \nn \\
    &=\frac{\mc{L}_\mc{F}}{16 \pi^2 r^4} \Biggl[
        Q_m^2 + \frac{\mc{L}_\mc{G} Q_m^2}{\mc{L}_\mc{F}^2} \biggl(
            - \mc{L}_\mc{G} - \theta_{Q_m} \mc{L}_\mc{G}+ \frac{\mc{L}_\mc{G} }{\mc{L}_\mc{F}} \,\theta_{Q_m} \mc{L}_\mc{F}
        \biggr)
    \Biggr]
    \nn \\
    & \quad 
    + \frac{\mc{L}_\mc{G}}{16 \pi^2 r^4 \mc{L}_\mc{F}}\Biggl[
        2 \mc{L}_\mc{G} Q_m^2 + Q_m^2 \,\theta_{Q_m}\mc{L}_\mc{G}- Q_m ^2\frac{\mc{L}_\mc{G}}{\mc{L}_\mc{F}} \,\theta_{Q_m}\mc{L}_\mc{F}
    \Biggr]
    \nn \\
    &= \frac{Q_m^2(\mc{L}_\mc{F}^2 + \mc{L}_\mc{G}^2)}{16 \pi^2 r^4 \mc{L}_\mc{F}}
    \,.
    \label{eq:delLdelQm-app}
\end{align}
It should be noted that the derivatives of $\mc{L}_\mc{F}$ and $\mc{L}_\mc{G}$ such as $\theta_{r}\mc{L}_\mc{F}$ are canceled in these calculations.
If we have an electric charge, the results are a bit complicated due to the non-vanishing contributions from the derivatives, because of which it becomes more tractable to bypass magnetic black holes to study dyonic black holes as discussed in section~\ref{sec:dyonic}.
Eqs.~\eqref{eq:delLdelr-app} and \eqref{eq:delLdelQm-app} lead to the following relations:
\begin{align}
    \theta_r \mc{L} = - 4 (\theta_\mc{F} + \theta_\mc{G}) \mc{L}\,,
    \quad 
    \theta_S \mc{L} = - 2 (\theta_\mc{F} + \theta_\mc{G}) \mc{L}\,,
    \quad 
    \theta_{Q_m} = 2 (\theta_\mc{F} + \theta_\mc{G}) \mc{L}\,,
\end{align}
where we used $S \propto r^2$.

\newcommand{\arxivfont}{\rmfamily}
\bibliographystyle{yautphysm}
\bibliography{ref}

\end{document}